\documentclass[%
 aip,
onecolumn��
 amsmath,amssymb,
 preprint,%
]{revtex4-1}

\usepackage{graphicx}
\usepackage{dcolumn}
\usepackage{bm}

\usepackage[utf8]{inputenc}
\usepackage[T1]{fontenc}
\usepackage{mathptmx}

\usepackage{array}

\usepackage{amsfonts}
\usepackage{amsmath}
\usepackage{multirow}
\usepackage{float}
\usepackage{tikz}
\usepackage{makecell}

\tikzstyle{mybox} = [draw=black, thick,
rectangle, rounded corners, inner sep=10pt, inner ysep=20pt]
\tikzstyle{fancytitle} =[fill=black, text=white]

\usepackage{booktabs,siunitx} 
\usepackage{multirow}

\begin{document}

\newtheorem{thm}{Theorem}[section]
\newtheorem{lemma}[thm]{Lemma}
\newtheorem{prop}[thm]{Proposition}
\newtheorem{rem}[thm]{Remark}
\newtheorem{cor}[thm]{Corollary}

\preprint{AIP/123-QED}

\title[Sample title]{When Machine Learning Meets Multiscale Modeling in Chemical Reactions}

\author{Wuyue Yang}
\altaffiliation[]{These authors have contributed equally to this work.}
\affiliation{Zhou Pei-Yuan Center for Applied Mathematics, Tsinghua, Beijing, 100084, P.R. China\\}%
\author{Liangrong Peng}%
\altaffiliation[]{These authors have contributed equally to this work.}
\affiliation{College of Mathematics and Data Science, Minjiang University, Fuzhou, 350108, P.R. China
}

\author{Yi Zhu}
\affiliation{Zhou Pei-Yuan Center for Applied Mathematics, Tsinghua, Beijing, 100084, P.R. China\\}%

\author{Liu Hong}
\email{zcamhl@tsinghua.edu.cn}
\affiliation{Zhou Pei-Yuan Center for Applied Mathematics, Tsinghua, Beijing, 100084, P.R. China\\}%

\date{\today}

\begin{abstract}
Due to the intrinsic complexity and nonlinearity of chemical reactions, direct applications of traditional machine learning algorithms may face with many difficulties. In this study, through two concrete examples with biological background, we illustrate how the key ideas of multiscale modeling can help to reduce the computational cost of machine learning a lot, as well as how machine learning algorithms perform model reduction automatically in a time-scale separated system. Our study highlights the necessity and effectiveness of an integration of machine learning algorithms and multiscale modeling during the study of chemical reactions.
\end{abstract}


\maketitle
\section{Introduction}
Accompanied with the matter synthesis and decomposition, energy storage and release, biofunction activation and deactivation, chemical reactions play a fundamental role in multiple disciplines \cite{Janos1989Mathematical}, including biology, chemical engineering, materials science and so on.  They help to model complicated phenomena in nature by an explicit reaction network, to allow the interpretation of observed data through quantitative mathematical equations, and to translate varied experimental conditions into tunable reaction rates and reaction orders. Due to their high complexity and nonlinearity, the previous studies of chemical reactions heavily rely on sophisticated mathematical analysis and first-principle calculations, like quantum chemistry \cite{Johnson2003Quantum}.

The first mission of studies on chemical reactions is to obtain the proper mathematical model which can interpret the observed phenomena and data. Even though there are some empirical laws and some pre-knowledge on the reaction networks which may help to build the model, the parameters like reaction rates are usually deeply buried inside the massive data.  Recent rapid development of various machine learning algorithms, especially deep neural networks, make inferring reaction networks and parameters be possible and efficient. Mangan \textit{et al.} \cite{mangan2016inferring} proposed an implicit sparse identification of nonlinear dynamics to infer hidden biochemical reaction networks, with emphasis on the rational nonlinear forms of the governing dynamics. Hu \textit{et al.} \cite{hu2020revealing} constructed a so-called ODENet (short for Ordinary Differential Equations Network), which was used for explicitly modeling the Lotka-Volterra type dynamics and actin growth in the presence of  medium-level noises. From a stochastic perspective, the chemical reaction system was modeled as a continuous-time Markov chain, whose propensity function was reconstructed as a combination of the pre-designed basis functions based on the maximization of log-likelihood function \cite{zhang2019learning}.

Costello and Martin \cite{costello2018machine} showed that a supervised learning method can predict the metabolic pathway dynamics from proteomics data, which may be used to design various bioengineered systems. Yang \textit{et al.} \cite{yang2019prediction} revealed that the aggregation rates of amyloid proteins could be reliably estimated based on the feedforward fully connected neural network and feature selection.
In organic chemistry, given some reactants and external conditions, all possible reactions were ranked by a machine learning approach, including a reactive site classifier and a ranking model, with the top-ranked mechanism corresponding to the major products \cite{kayala2011learning}.
The Gaussian process regression was utilized to construct the potential energy surface of the HO$_x$ system \cite{song2020revisiting}, which could reduce the computational cost and meanwhile guarantee the convergence with fewer training points.
For more applications of machine learning to chemical reactions, including the supervised and unsupervised learning, see \textit{e.g.} Refs. \cite{villaverde2014reverse, brunton2016discovering, boninsegna2018sparse, choi2018robust, daniels2015automated}. 

The successful attempts of machine-learning-based modeling pave a new way to understand the complicated dynamics of chemical reactions. However, most chemical reactions involve plenty of reactants, multiple potential reaction routines, diverse reaction rates and so on. Without considering the this intrinsic multi-component and multiscale nature of the system, direct applications of machine learning algorithms may face inevitable difficulties (see examples below for details). Motivated by the requirements on a real complex system, especially a simultaneous maintenance of the efficiency of macroscopic models and the accuracy of microscopic models, the view of multiscale modeling is introduced. It focuses on a proper separation of the system or phenomenon into several scales with minimum overlap, a correct characterization of the relation between different levels of physical models, as well as a systematical procedure of coarse-graining \cite{Weinan2011Principles}. Multiscale modeling offers a unified way to examine the system of chemical reactions, by looking into the reactions occurring at different time scales and the relations between them. Therefore, it is expected that a proper integration of machine learning algorithms with ideas and methodology of multiscale modeling and analysis will shed some light into this field. And this leads to the major motivation of our current study.

To be concrete, we will justify our arguments from two aspects: (1) By using the explicit correspondence between mesoscopic chemical master equations and macroscopic mass-action equations in Kurtz's limit, the challenging task of learning detailed probability distribution function (PDF) is converted into learning low-order moments. Obviously, the latter is much easier. In this case, the computational cost of direct machine learning is greatly reduced by incorporating the multiscale modeling. (2) When fast and slow reactions appear simultaneously in the same system, meaning there is a time-scale separation among the chemical reactions, the ODENet -- a kind of machine learning algorithms with sparse identification show an astonishing ability of deriving simplified models under Quasi Steady State Approximation (QSSA) automatically. Therefore, machine learning could help to model multiscale chemical reactions too. These two examples clearly demonstrate that machine learning and multiscale modeling are closely related to each other. A proper integration of two approaches will greatly facilitate our study of chemical reactions.

The whole paper is organized as follows. A basic architecture of the ODENet, a special kind of machine learning algorithms which is designed to derive the explicit form of ODEs from the pre-given time series data, is introduced in Section II. Along with the basic ideas and techniques for multiscale modeling and analysis for chemical reactions, including the Kurtz's limit from chemical master equations to mass-action equations, and the quasi steady-state approximation. In Section III, we illustrate our key ideas through two examples -- the development and differentiation of cells, as well as the self-regulatory gene transcription and translation. The usefulness of an integration of machine learning and multiscale modeling could be clear learned. The last section contains some discussions.

\begin{figure}[]
	\centering
	\includegraphics[width=1.0\linewidth]{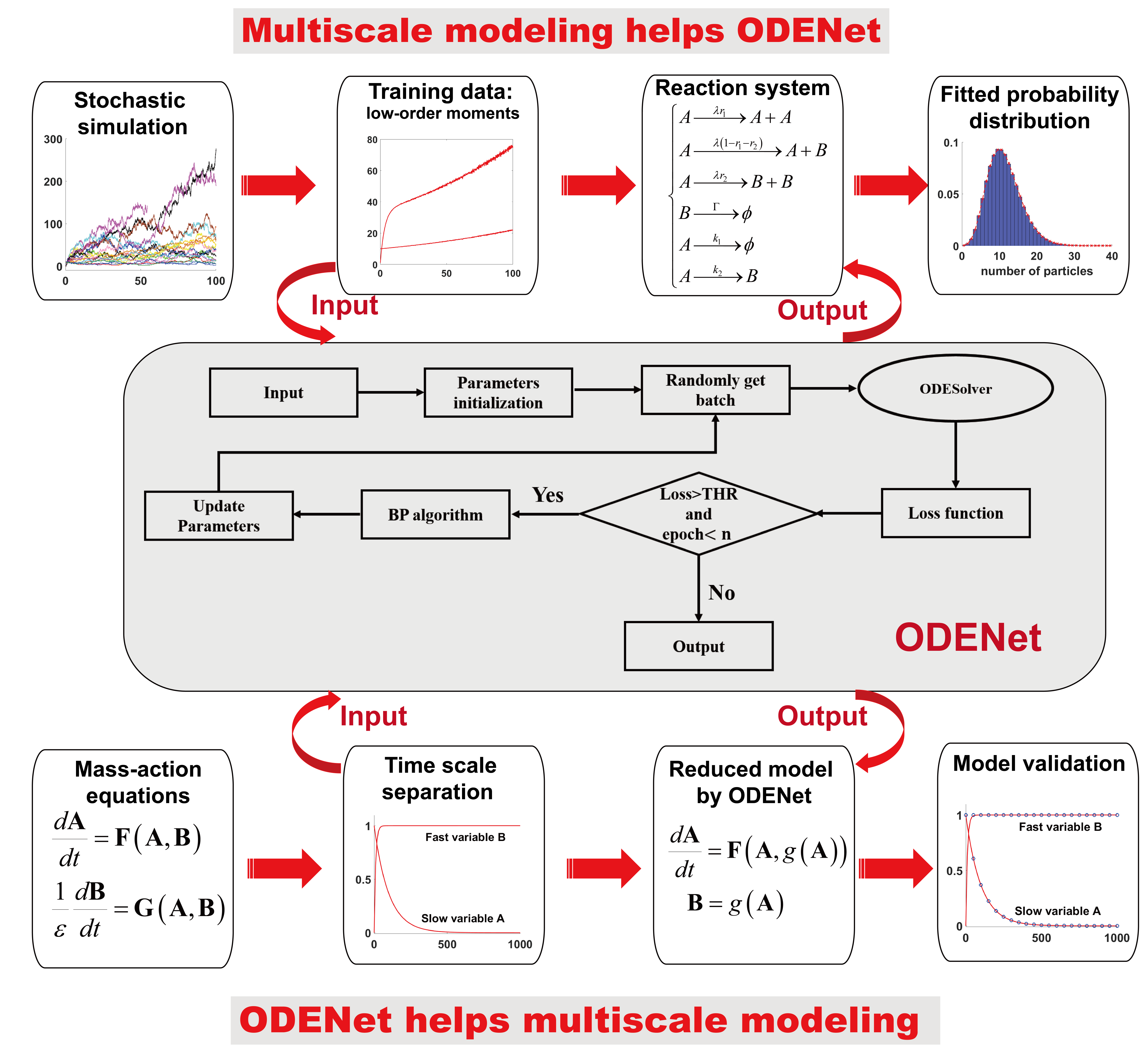}
	\caption{\textbf{An integration of ODENet with multi-scale modeling in the study of chemical reactions.} The upper panel illustrates the ODENet-based learning procedure of reaction mechanism under the help of multiscale modeling, while the lower panel gives the automatic procedure for model reduction aided by ODENet. The flowchart of ODENet is shown in the middle.}
	\label{fig.flowchart}
\end{figure}

\section{Methods}

\subsection{Basic Architecture of ODENet}

The ordinary differential equations network was proposed \cite{chen2018neural, hu2020revealing} as a continuous version of the famous ResNet \cite{he2016deep} for dealing with time series data modeled by ordinary differential equations (ODEs). Mathematically, the consecutively repeating building blocks -- each layer of a residual network can be expressed as $\mathbf{y}_{k+1}=\mathbf{y}_{k}+f\left(\mathbf{y}_{k}; \theta_{k}\right)$, where $\mathbf{y}_{k}$ is the output of $k^{th}$ hidden layer, $\mathbf{y}_{k+1}$ is the output of $(k+1)^{th}$ hidden layer and $f\left(\mathbf{y}_{k}; \theta_{k}\right)$ represents the function of a network layer parameterized by $\theta_k$. After a simple algebraic transformation, we can get $\frac{\mathbf{y}_{k+1}-\mathbf{y}_{k}}{h}=\frac{f\left(\mathbf{y}_{k}; \theta_{k}\right)}{h}$, which is the Euler's discretization scheme of ODEs,
\begin{equation}
\frac{d \mathbf{y}}{d t}=\frac{f(\mathbf{y} ; \theta)}{h}.
\end{equation}
As a consequence, the forward propagation process of a residue network is actually equivalent to the numerical solvation of a group of corresponding ordinary differential equations. Alternatively, it also means if we use an ODE solver to solve the ODEs directly, the process of forward propagation in a residue network is accomplished too. This significant finding lays down the theoretical foundation of ODENet. The application of ODE solvers could easily cope with input data with unequal time intervals, fight against medium-level noises, control the numerical errors and dynamically adjust its convergence criteria.

To enhance the ability of learning the explicit governing ODEs from the pre-given time series data, in a previous work we combined the ODENet with symbolic regression and sparse identification \cite{hu2020revealing}. Symbolic regression means the explicit form of $f\left( {y;\theta } \right)$ is characterized through parameters $\theta$ by expanding $f(y)$ on a complete set of orthogonal basis functions $\Gamma(y)$, \textit{i.e.} $f\left( {y;\theta } \right)=\theta\Gamma(y)$. Consequently, the learning of ODEs becomes to determine the unknown parameters $\theta$ from the data. In practice, polynomials are the most often used basis functions. Sparse identification means in the loss function $L$, an additional regulation term $\|\theta\|_{1}$ is added in order to remove redundant free parameters $\theta$ as many as possible. So that the loss function contains two parts:
\begin{equation}
L=\|\mathbf{y}-\widehat{\mathbf{y}}\|_{1}+\varepsilon\|\theta\|_{1}.
\end{equation}
The first part controls the difference between the training data $\mathbf{y}$ and the predicted data $\widehat{\mathbf{y}}$ by ODENet, while the second part aims at a minimal model according to the Occam's razor. Here $\varepsilon$ is a hyperparameter. To obtain the optimal parameters $\theta$, the classical Back Propagation (BP) algorithm \cite{lecun1989backpropagation} is adopted to make an update, which will be repeated for many iterations until the loss function converges or is less than the threshold. Please see Fig. \ref{fig.flowchart} on the flowchart of ODENet or refer to Ref.\cite{hu2020revealing} for further details.

\subsection{Multiscale Modeling of Chemical Reactions}
Without loss of generality, we consider a chemical system with $N$ species and $M$ reactions \cite{othmer2003},
\begin{equation}
\label{cr}
\nu_{1j}S_1+\nu_{2j}S_2+\cdots+\nu_{Nj}S_N \stackrel{{k_j}}{\longrightarrow} \nu^{'}_{1j}S_1+\nu^{'}_{2j}S_2+\cdots+\nu^{'}_{Nj}S_N, \quad
j=1,2,\cdots,M,
\end{equation}
where $k_j>0$ denotes the rate constant of the reaction $j$.
The nonnegative integers $\{\nu_{ij}\}$  and $\{\nu^{'}_{ij}\}$ denote the stoichiometric coefficients of the reactants and products respectively. The stoichiometric matrix is introduced as $U=[(u_{ij})]_{N \times M}$ with elements being $u_{ij}=\nu^{'}_{ij}-\nu_{ij}$.

\subsubsection{Chemical Master Equations}
We focus on the molecular number of species $(S_1,S_2,\cdots, S_N)$ represented by a stochastic variable $\bold{n}=(n_1,n_2, \cdots,n_N)^T$ in a reaction vessel of volume $V$.
When the magnitude of $(n_1,n_2, \cdots,n_N)^T$ is relatively small compared with the Avogadro's constant, the randomness comes into play due to the intrinsic stochasticity of molecular collisions. From the perspective of ensemble average, we can denote the probability of the system in the state $\bold{n}$ by $p(\bold{n},t)$, where the time-dependence is usually omitted as $p(\bold{n})$.

With respect to the reactions in \eqref{cr}, the probability distribution obeys the following chemical master equations (CMEs), in a compact form as,
\begin{equation}\label{cme}
\frac{d}{dt}p(\bold{n})
=\sum_{j=1}^{M} \left[p(\bold{n}-\bold{u}_j) \Phi_j(\bold{n}-\bold{u}_j) - p(\bold{n}) \Phi_j(\bold{n}) \right],
\end{equation}
accompanied by the initial condition $p(\bold{n})|_{t=0}=p_0(\bold{n})$.
Here $\bold{u}_j$ is the $j$-th column of stoichiometric matrix $U=(\bold{u}_1, \bold{u}_2, \cdots, \bold{u}_M)$, and $\Phi_j(\bold{n})$ is the mesoscopic propensity function characterizing the probability $\Phi_j(\bold{n}) dt$ for which the $j-$th reaction occurs once within the time interval $[t, t+dt)$.

In general, the state-dependent mesoscopic propensity function $\Phi_j(\bold{n})$ of CMEs is assumed to follow the laws of mass-action,
\begin{equation}\label{rate}
\Phi_j(\bold{n})
=k_j V \prod_{l=1}^{N} \left(V^{-\nu_{lj}} C_{n_l}^{\nu_{lj}} \right),
\end{equation}
which is the product of molecular number in a polynomial form and the rate coefficient $k_j$ for ${n_l}\geq \nu_{lj}, \forall l=1,2,\cdots, N$. When there are not enough particles to form a reactant, saying $S_l$, such that ${n_l} < \nu_{lj}$, the propensity reduces to zero, $\Phi_j(\bold{n})=0$.

\subsubsection{Stochastic Simulations}
In most cases, the chemical master equations in (\ref{cme}) are a huge group of ordinary differential equations, which are quite computational consuming. Alternative efficient sampling algorithms are needed. The Gillespie algorithm (GA) \cite{gillespie1977exact}, which is able to generate typical time evolutionary trajectories of species according to the reaction mechanisms and reaction rates in a stochastic way, maybe the most famous one.

Gillespie implemented two stochastic simulation algorithms. The one is the direct method (DM) and the other is the first-reaction method (FRM). These two methods are theoretically equivalent, so we here only implement the first reaction method. The FRM generates putative time for every reaction and chooses a time at which the corresponding reaction would occur while no other reaction occurred before that.

By independently running the Gillespie algorithm once and again, statistics on the corresponding stochastic trajectories will converge to the corrected probability distribution given by the chemical master equations. They constitute the training data set to feed into the machine learning algorithms.

\subsubsection{Moment-Closure Equations in Kurtz's Limit}
Although CMEs provide a relatively accurate way to model general chemical reaction systems, it leads to a heavy burden in both modeling and experiments since the dimensionality of the transition matrix is usually extremely high. Moreover, the time-consuming numerical simulation of CMEs becomes a common bottleneck when the number of species or reactions is large. In order to make a simplification, we turn to look at the mean density of species,
\begin{equation}\label{mean_variance}
c_i=\sum_{\bold n} V^{-1}n_i p(\bold n),
\end{equation}
when the molecular number of reactants becomes large.

To deduce the macroscopic kinetics of the concentration $c_i$, we multiply \eqref{cme} by the number density $V^{-1}n_i$ and take the summation over all admissible state $\{\bold n\}$ on both sides, which yields,
\begin{equation}\label{mae1}
\begin{split}
\frac{d}{d t} \sum_{\bold n} V^{-1} n_{i} p(\bold n)
=&\sum_{j=1}^{M} \sum_{\bold n} V^{-1} n_{i}
\left[p(\bold{n}-\bold{u}_j) \Phi_j(\bold{n}-\bold{u}_j) - p(\bold{n}) \Phi_j(\bold{n}) \right]\\
=&\sum_{j=1}^{M}  u_{ij}
\left[\sum_{\bold n} V^{-1} p(\bold{n}) \Phi_j(\bold{n}) \right]
,
\end{split}
\end{equation}
where in the last step we have used the variable substitution $\bold{n}-\bold{u}_j=\bold{n}'$ and have neglected the boundary terms.
Direct calculation shows that the volume density of mesoscopic propensity function deduces, $V^{-1}\Phi_j(\bold{n})=\phi_j(V^{-1}\bold{n})+\mathcal{O}(V^{-1})$, with $\phi_j(\bold{c})=k_j \prod_{l=1}^N {{c_l}^{\nu_{lj}}}/{\nu_{lj}!}$ being the usual macroscopic propensity function.

Taking the limit of $V\rightarrow +\infty, \bold{n} \rightarrow +\infty$ while keeping $V^{-1} \bold n$ finite, we have the following mass-action equations (MAEs)
\begin{equation}\label{mae}
\frac{d}{dt}c_i(t)=\sum_{j=1}^{M}  (\nu^{'}_{ij}-\nu_{ij}) \phi_j(\bold c),
\end{equation}
on a nonnegative continuous state space $\{\bold c| \bold c \in \mathbb{R}^N_{\geq 0}\}$.
The MAEs in \eqref{mae} is the macroscopic description derived from the mesoscopic CMEs of the reaction system \eqref{cr}. A rigorous mathematical justification of the above limit process was first done by Kurtz in the 1970s \cite{kurtz1972}. Similar procedure can be carried out for high-order moments of PDF, like the second-order variance studied in the first example in Section III.

\begin{rem}
	According to the results proved by Kurtz \cite{kurtz1972}, in the limit of $V\rightarrow+\infty$, for any finite time the solution of CMEs in \eqref{cme} will converge in probability to the solution of the corresponding MAEs in \eqref{mae}, provided the initial conditions $\lim_{V\rightarrow+\infty} V^{-1} \bold n(t=0) = \bold c(t=0)$, which is a straightforward consequence of the Central Limit Theorem.
	Our derivation above from the CMEs in \eqref{cme} to MAEs in \eqref{mae} for the reaction system \eqref{cr} serves as a formal illustration of Kurtz's theorem.
\end{rem}

\subsection{Model Reduction by QSSA}
Consider a very general chemical reaction system with time scale separation, which is written in an abstract matrix form,
\begin{equation}
\left\{
\begin{split}
\frac{d\bold{A}}{dt}&=\bold{F}(\bold{A,B}),\\
\frac{1}{\epsilon}\frac{d\bold{B}}{dt}&=\bold{G}(\bold{A,B}),
\end{split}
\right.
\end{equation}
where $\bold{A}$ and $\bold{B}$ respectively stand for slow and fast variables after some kind of proper non-dimensionalization. $\epsilon\ll1$ is a small parameter characterizing the gap between fast and slow time scales in the dynamics.

With respect to above dynamics, QSSA states that in the slow time scale dominated by the changes in $\bold{A}$, $\bold{B}$ can be regarded as remaining at a dynamically equilibrium state (quasi steady state) due to their fast reactive nature, meaning approximately we have $\bold{G}(\bold{A,B})=0$. If $\bold{B}$ can be uniquely solved from this algebraic relation, \textit{i.e.} $\bold{B}=\bold{g}(\bold{A})$, the original time-scale separated dynamics could be simplified as
\begin{equation}
\frac{d\bold{A}}{dt}=\bold{F}(\bold{A,\bold{g}(\bold{A})}),\quad \bold{B}=\bold{g}(\bold{A}).
\end{equation}
QSSA is a very classical model reduction approach and has been widely used in the study of chemical reactions, see \textit{e.g.} Ref.\cite{Segel1989The} for details.

\section{Results and Discussion}
In this section, through two concrete examples -- the single proliferative compartment model (SPCM) of IFE (interfollicular epidermis) maintenance as well as a gene network with autoregulatory negative feedback, we are going to show how machine learning and multiscale modeling help each other in the study of chemical reactions.

\subsection{Single Proliferative Compartment Model}
\subsubsection{The Basic Model}
In the first example, the SPCM of IFE maintenance considered by Clayton \textit{et al.} \cite{clayton2007single} is adopted to illustrate how multiscale modeling helps to reduce the computational cost of machine learning during inferring the detailed reaction mechanisms and reaction rates. According to the observations by Clayton \textit{et al.} \cite{clayton2007single}, the clone fate of proliferating epidermal progenitor cells (EPCs) plays an essential role in adult epidermal homeostasis. And the key clone size distribution is modeled by chemical master equations, whose explicit forms are the major goal of machine learning. By taking the explicit correspondence between mesoscopic chemical master equations and macroscopic mass-action equations in the Kurtz's limit, the challenging task of learning detailed probability distribution function is converted into learning low-order moments. Obviously, the latter is much easier. A similar idea has been previously applied by one of the authors to investigate the kinetics of amyloid aggregation, but without referring to machine learning \cite{Hong2013Simple, Tan2013Modeling}.

\begin{figure}[H]
	\centering
	\includegraphics[width=1.0\linewidth]{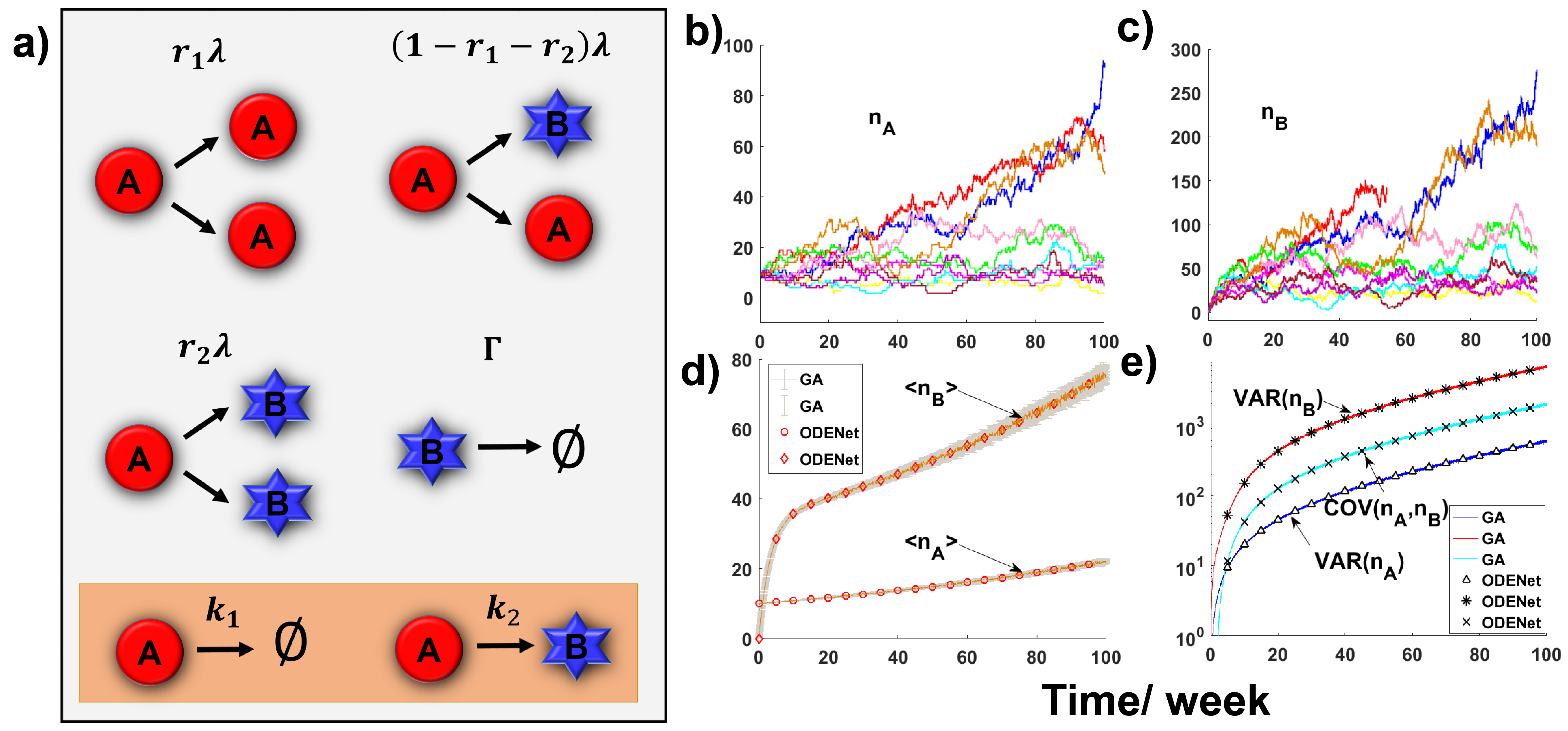}
	\caption{\textbf{Single proliferative compartment model.} (a)The mechanism of single proliferative compartment model. EPCs (red circles) have an unlimited self-division potential to maintain the epidermis at a rate of $r_{1} \lambda$. Proliferating EPCs cells divide into two post mitotic basal cells (blue stars) at a rate of $r_{2} \lambda$. Asymmetric divisions of EPCs into itself and post mitotic basal cells are at a rate of $1-(r_{1}+r_{2}) \lambda$. After mitosis in the basal layer, the post mitotic basal cells leak at a rate of $\Gamma$. $k_1$ and $k_2$ represent the rate constants for two additional possible reactions inferred by the ODENet. 10 typical stochastic trajectories for (b) $n_A$ and (c) $n_B$ are generated by GA. The learned results of ODENet are compared with the training data generated by GA on the (d) average and (e) variance of cell numbers. Here the rate constants in SPCM are set as $\lambda  = 1.1, {r_1} = 0.0836, {r_2} = 0.0764, \Gamma  = 0.31$ per week in accordance with \cite{clayton2007single}. The initial PDF is taken as a delta distribution with $p_0(10,0)=1$.}
	\label{fig.Gillespie}
\end{figure}

Consider two reactant species in the single-proliferative compartment model, including proliferating EPCs (denoted as $A$) and post-mitotic cells in the basal layer ($B$). There are four reactions which involve symmetric cell division and  asymmetric cell division. As shown in Fig. \ref{fig.Gillespie}a, $A$ has a unlimited self-reproduction potential at a rate $r_{1}\lambda$ in order to maintain the epidermis, where $\lambda$ is the integrated division rate of proliferating EPCs $A$. $A$ can also differentiate into $A + B$ or $B + B$ at the rate of $(1-r_{1}-r_{2})\lambda$ or $r_{2}\lambda$, respectively. In the transfer process, $B$ cells in the basal layer leak from the clone-size distributions at $\Gamma$ rate. The above reaction system reduces to the one studied by Clayton \textit{et al.} \cite{clayton2007single} with symmetric division rates $r_1=r_2$.

With respect to the SPCM and coefficients given in Fig. \ref{fig.Gillespie}, $10^6$ times independent stochastic simulations are performed by using the Gillespie algorithm. They constitute the training data set to feed into our following ODENet-based machine learning procedure.


\subsubsection{Learning Mass-Action Equations by ODENet}

Here our major goal is to obtain the SPCM in Fig. \ref{fig.Gillespie}a and the explicit rate constants. However, a direct application of ODENet to learn the time evolution of $p(n_A,n_B)$ (or the chemical master equations) from the training data generated by stochastic simulations is prohibited due to heavy computational cost. Therefore, by taking advantage of the knowledge of multiscale modeling in chemical reactions, especially the explicit correspondence between mesoscopic chemical master equations and macroscopic mass-action equations in the Kurtz's limit, we turn to learn low-order moments instead of the probability distribution function governed by chemical mass-action equations.

With respect to training data of $\left\langle n_{A}\right\rangle=\sum n_{A} p\left(n_{A}, n_{B}\right)$ and $\left\langle n_{B}\right\rangle=\sum n_{B} p\left(n_{A}, n_{B}\right)$ by averaging the stochastic trajectories generated through Gillespie algorithms, we need to determine the exact types of chemical reactions involving with these two reactants, their reaction orders and reaction rate constants. Without loss of generality, here we make a cutoff on the chemical reactions up to the second order, corresponding to a combination of $A, B, A + A, A + B, B + B$, which reads
\begin{equation}
	\left\{\begin{array}{l}
		\displaystyle\frac{d \left\langle n_{A}\right\rangle}{d t}=\alpha_{11} \left\langle n_{A}\right\rangle+\alpha_{12} \left\langle n_{B}\right\rangle+\alpha_{13} \left\langle n_{A}\right\rangle^{2}+\alpha_{14} \left\langle n_{A}\right\rangle \left\langle n_{B}\right\rangle+\alpha_{15} \left\langle n_{B}\right\rangle^{2}, \\
		\displaystyle\frac{d \left\langle n_{B}\right\rangle}{d t}=\alpha_{21} \left\langle n_{A}\right\rangle+\alpha_{22} \left\langle n_{B}\right\rangle+\alpha_{23} \left\langle n_{A}\right\rangle^{2}+\alpha_{24} \left\langle n_{A}\right\rangle \left\langle n_{B}\right\rangle+\alpha_{25} \left\langle n_{B}\right\rangle^{2}.
	\end{array}\right.
	\label{crt}
\end{equation}

Now we implement the ODENet to learn the dynamics in (\ref{crt}). Clearly, not all reaction rate constants will appear in the final model. Those redundant coefficients will be picked out by ODENet and removed through sparse identification. After training and regression, only three non-zero coefficients $\alpha_{11}=0.0079$, $\alpha_{21}=1.0903$ and $\alpha_{22}=-0.3094$ are kept in the final results.


\subsubsection{Learning High-Order Moment Equations}

During the learning procedure of ODENet, since all coefficients in front of quadratic terms in (\ref{crt}) are removed, we can make a conclusion that only first-order reactions are present in the current system. Then with respect to above learned dynamics and coefficients, the desired single proliferative compartment model as shown in Fig. \ref{fig.Gillespie}a is reconstructed by ODENet. However, the existence of two additional reactions, $A \stackrel{k_{1}}{\longrightarrow}{\phi}$ and $A \stackrel{k_{2}}{\longrightarrow} B$ (see orange box in Fig. \ref{fig.Gillespie}a) could not be excluded in principle, which means at the moment the probability distribution of $A$-type and $B$-type cells follows
\begin{equation}
\begin{aligned}
\frac{d}{d t} p\left(n_{A}, n_{B}\right)=& k_{1}\left[\left(n_{A}+1\right) p\left(n_{A}+1, n_{B}\right)-n_{A} p\left(n_{A}, n_{B}\right)\right] \\
&+k_{2}\left[\left(n_{A}+1\right) p\left(n_{A}+1, n_{B}-1\right)-n_{A} p\left(n_{A}, n_{B}\right)\right] \\
&+\lambda \big[
r_1(n_A-1) p(n_A-1,n_B) + r_2(n_A+1) p(n_A+1,n_B-2) \\
&+ (1-r_1-r_2)n_A p(n_A,n_B-1)-n_Ap(n_A,n_B) \big]\\
&+\Gamma \big[(n_B+1) p(n_A,n_B+1)-n_Bp(n_A,n_B) \big].
\end{aligned}
\label{mastereq_extension}
\end{equation}
Here the same notations are borrowed just for simplicity. It should be noted that at the moment we still have no precise knowledge on all six  reaction rate constants $k_1$, $k_2$, ${r_1}$, ${r_2}$, $\lambda $ and $\Gamma$.

To determine the unknown coefficients, we further go to the second-order of PDF, the variance of cell numbers to be exact. Based on (\ref{mastereq_extension}), the first-order (average) and second-order moments (variance) of $n_A$ and $n_B$ evolve according to
\begin{equation}
\left\{\begin{array}{l}
\vspace{2ex}
\displaystyle\frac{d\left\langle n_{A}\right\rangle}{d t}=\left[\left(r_{1}-r_{2}\right) \lambda-k_{1}-k_{2}\right]\left\langle n_{A}\right\rangle,\\
\vspace{2ex}
\displaystyle\frac{d\left\langle n_{B}\right\rangle}{d t}=\left[k_{2}+\left(1-r_{1}+r_{2}\right) \lambda\right]\left\langle n_{A}\right\rangle-\Gamma\left\langle n_{B}\right\rangle,\\
\vspace{2ex}
\displaystyle\frac{d V_{A}}{d t}=\beta_{11}\left\langle n_{A}\right\rangle+\beta_{12} V_{A}, \\
\vspace{2ex}
\displaystyle\frac{d V_{B}}{d t}=\beta_{21}\left\langle n_{A}\right\rangle+\beta_{22}\left\langle n_{B}\right\rangle+\beta_{23} V_{B}+\beta_{24} C o v, \\
\vspace{2ex}
\displaystyle\frac{d C o v}{d t}=\beta_{31}\left\langle n_{A}\right\rangle+\beta_{32} V_{A}+\beta_{33} C o v,
\end{array}\right.
\label{variance}
\end{equation}
where $V_{A}=\left\langle n_{A}^{2}\right\rangle-\left\langle n_{A}\right\rangle^{2}$, $V_{B}=\left\langle n_{B}^{2}\right\rangle-\left\langle n_{B}\right\rangle^{2}$ and $Cov=\left\langle n_{A}n_{B}\right\rangle-\left\langle n_{A}\right\rangle\left\langle n_{B}\right\rangle$. Furthermore, we have $\beta_{11}=k_{1}+k_{2}+\left(r_{1}+r_{2}\right) \lambda$, $\beta_{12}=-2\left(k_{1}+k_{2}\right)+2\left(r_{1}-r_{2}\right) \lambda$, $\beta_{21}=k_{2}+\left(1-r_{1}+3 r_{2}\right) \lambda$, $\beta_{22}=\Gamma$, $\beta_{23}=-2 \Gamma$, $\beta_{24}=2\left(\left(1-r_{1}+r_{2}\right) \lambda+k_{2}\right)$, $\beta_{31}=-\left(2 r_{2} \lambda+k_{2}\right)$, $\beta_{32}=\left(1-r_{1}+r_{2}\right) \lambda+k_{2}$, $\beta_{33}=\left(r_{1}-r_{2}\right) \lambda-k_{1}-k_{2}-\Gamma$. Now following the same procedure, the unknown values of $\beta$'s could be learned by ODENet and are summarized in SI.


\subsubsection{Deriving Chemical Master equations}

The relations among desired rate constants $k_1$, $k_2$, $r_1$, $r_2$, $\lambda$, $\Gamma$ and those learned parameters $\alpha$'s and $\beta$'s are stated through the following matrix, \textit{i.e.}
\begin{equation}
\underbrace{
\left[\begin{array}{cccccc}
1 & -1 & -1 & -1 & 0 & 0 \\
-1 & 1 & 0 & 1 & 1 & 0 \\
0 & 0 & 0 & 0 & 0 & 1 \\
1 & 1 & 1 & 1 & 0 & 0 \\
2 & -2 & -2 & -2 & 0 & 0 \\
-1 & 3 & 0 & 1 & 1 & 0 \\
0 & 0 & 0 & 0 & 0 & 1 \\
0 & 2 & 0 & 1 & 0 & 0 \\
-1 & 1 & 0 & 1 & 1 & 0 \\
1 & -1 & -1 & -1 & 0 & -1
\end{array}\right]
}_{\text{$V$}}
\underbrace{
\left[\begin{array}{c}
{r_1}\lambda  \\
{r_2}\lambda  \\
k_{1} \\
k_{2} \\
\lambda \\
\Gamma
\end{array}\right]
}_{\text{${\hat{u}}$}}
=
\underbrace{
\left[\begin{array}{c}
{\alpha _{11}} \\
{\alpha _{21}} \\
{-\alpha _{22}} \\
{\beta _{11}} \\
{\beta _{12}} \\
{\beta _{21}} \\
{\beta _{22}} \\
{-\beta _{31}}  \\
{\beta _{32}} \\
{\beta _{33}}
\end{array}\right]
}_{\text{{${\hat{b}}$}}}
.
\label{Overdetermined}
\end{equation}
Direct calculations show that the rank of the augmented matrix ($rank(V| \hat b)= 7$) is larger than that of the coefficient matrix ($rank(V)=6$), meaning the linear equations in (\ref{Overdetermined}) constitute an overdetermined system, which can be solved through the Least Square Method. The unique least-square solution is given by $\hat{u}=\left(V^{T} V\right)^{-1} V^{T} \hat{b}$, whose relative errors with respect to the true values are less than $8\%$.

\begin{table}[H]
	\setlength\extrarowheight{2pt}
	\setlength{\tabcolsep}{9pt}
    \centering
	\begin{tabular}{cccccccc}
		\Xhline{1.2pt}
		\hline
		\multicolumn{7}{c}{Parameters}                                                                        \\ \hline
		& $k_1$        & $k_2$        & $r_1$       & $r_2$  & $\lambda$ & $\Gamma$ \\ \hline
		true value      & 0           & 0            & 0.0836 & 0.0764 & 1.1       & 0.31     \\ \hline
		learned value      & 0.0123       & -0.0197      & 0.0831 & 0.0824 & 1.1059    & 0.3074   \\ \hline
		relative errors & $\sim$       & $\sim$       & 0.60\% & 7.85\% & 0.54\%    & 0.84\%   \\ \hline
		\Xhline{1.2pt}
	\end{tabular}
	\caption{Comparison on the learned rate constants for (\ref{mastereq_extension}) by ODENet with the true values.}
	\label{SPCM_ODENet}
\end{table}

\begin{figure}[H]
	\centering
	\includegraphics[width=1.0\linewidth]{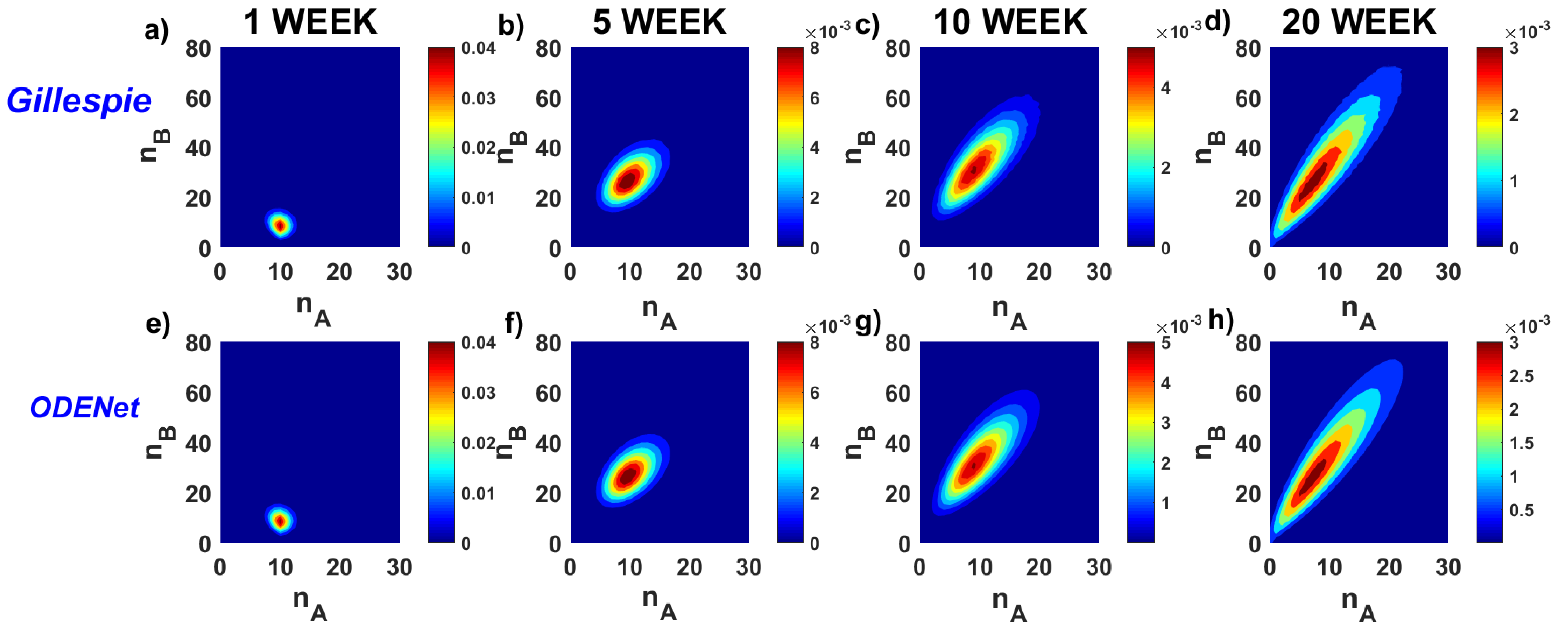}
	\caption{\textbf{Comparison of PDF generated by GA with the learned results of ODENet.} Joint probability distributions $p(n_A, n_B)$ are shown in (a,e) 1, (b,f) 5, (c,g) 10 and (d,h) 20 weeks respectively.}
	\label{fig.marginal}
\end{figure}

Even though $k_1$ and $k_2$ are not exactly identified as zero, their values are about one order of magnitude smaller than the others. In this sense, we have successful reconstructed the original SPCM based on the stochastic time trajectories of $n_A$ and $n_B$ in the training data set. As further validated in Fig. \ref{fig.marginal}, the joint probability distribution of the desired single-proliferative compartment model is honestly reproduced (see SI for the marginal probability distribution), which highlights the efficiency and effectiveness of our integrated approach of ODENet with multiscale modeling during the study of chemical reactions.

\subsection{A Gene Network with Autoregulatory Negative Feedback}

\subsubsection{The Basic Model}
\begin{figure}[H]
	\centering
	\includegraphics[width=0.9\linewidth]{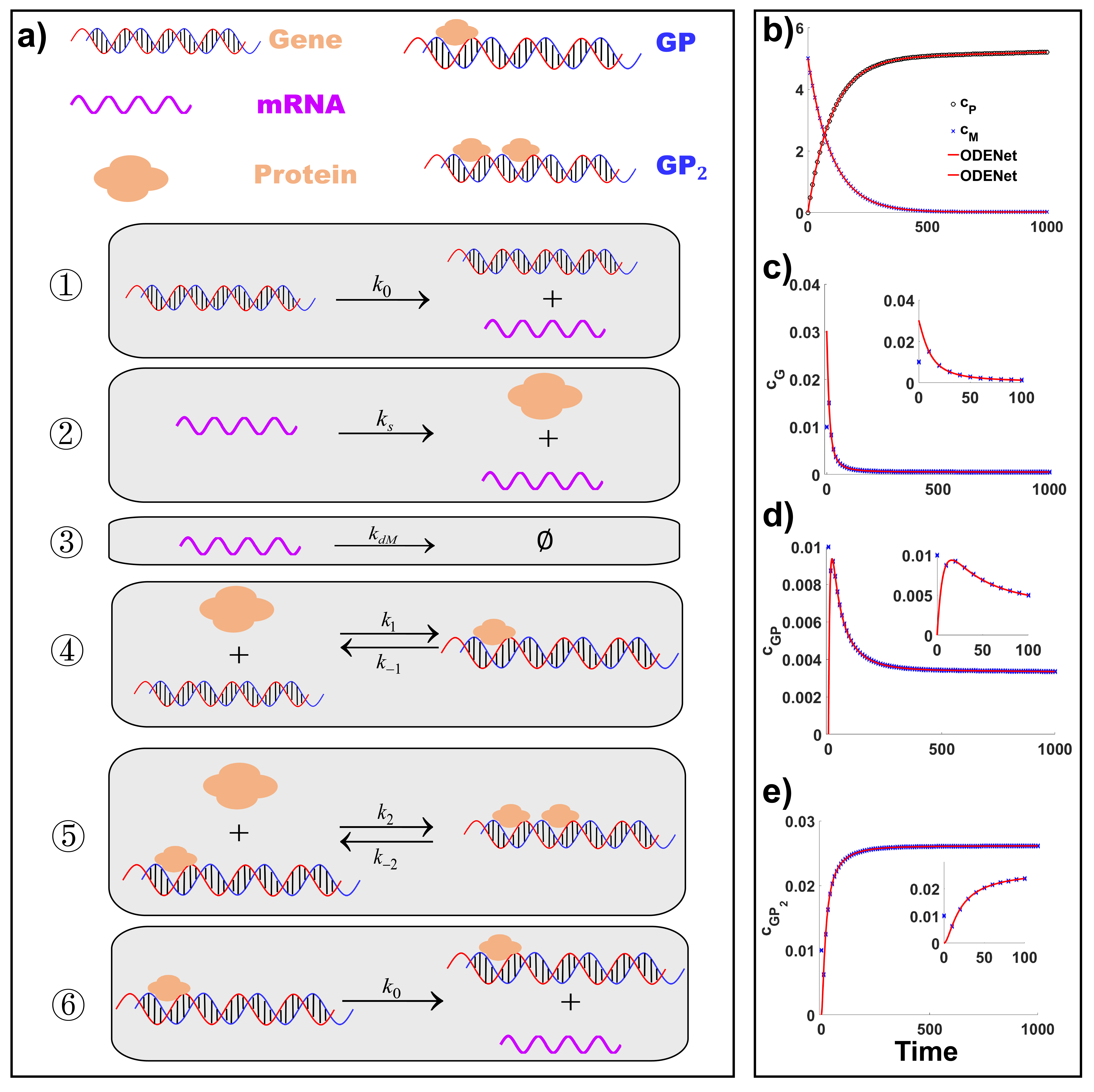}
	\caption{\textbf{Validation of ODENet aided model reduction.} (a) A cartoon illustration of the gene network with negative feedback, including transcription, translation, degradation and a negative feedback loop. Predictions of the reduced model in (\ref{qssa2}) (blue crosses) are compared with the original model in (\ref{mae4}) (red solid lines) on concentrations of (b) protein and mRNA in the slow time scale, and (c) gene and (d-e) gene-protein complexes in the fast time scale.}
	\label{fig.qssa}
\end{figure}

In the second example, we plan to show how machine learning can be used for model reduction, an important aspect of multiscale modeling with vast applications in chemical reactions. To illustrate our ideas, let us consider a gene network with autoregulatory negative feedback, which includes five reactants -- the gene ($G$), mRNA ($M$), protein ($P$), and two gene-protein complexes ($GP, GP_2$). Among them, there are eight reactions (see Fig. \ref{fig.qssa}a). $k_0, k_s, k_{dm}$ are rate constants of transcription from the gene $G$, translation into the protein $P$, and mRNA degradation, respectively. The gene can bind with either one or two proteins, whose forward and backward reaction rate constants are denoted as $k_1$, $k_{-1}$, $k_2$ and $k_{-2}$ separately. Furthermore, it is assumed that $GP$ produces mRNA at the same rate $k_0$ as the transcription rate of $G$ alone.

Macroscopically, the gene network in Fig. \ref{fig.qssa}a is described by chemical mass-action equations,
\begin{equation}\label{mae4}
\left\{
\begin{split}
&\frac{d}{dt}c_P= k_s c_{M} -k_1 c_G c_P + k_{-1}c_{GP}-k_2c_{P} c_{GP} + k_{-2} c_{GP_2},\\
&\frac{d}{dt}c_{M}=k_0 c_G + k_0 c_{GP} - k_{dM} c_{M},\\
&\frac{d}{dt}c_G= -k_1 c_G c_P + k_{-1}c_{GP},\\
&\frac{d}{dt}c_{GP}= k_1 c_G c_P - k_{-1}c_{GP}-k_2c_{P} c_{GP} + k_{-2} c_{GP_2},\\
&\frac{d}{dt}c_{GP_2}=k_2c_{P} c_{GP} - k_{-2} c_{GP_2}.
\end{split}
\right.
\end{equation}
It is noted that the total gene concentration is a constant due to the conservation law, \textit{i.e.} $c_{G}+c_{GP}+c_{GP_2}=c_{total}$. To produce a time-scale separation of reactions, we choose
$k_1=3, k_{-1}=2.4$, $k_2=9, k_{-2}=6$, $k_0=0.05, k_s=0.01, k_{dm}=0.01$, meaning the concentrations of $G, GP, GP_2$ can quickly reach dynamical balance in comparison with those of mRNA and protein. The initial conditions are set as
$c_G=c_{GP}=c_{GP_2}=0.01$, $c_P=0$, $c_M=5$.

\subsubsection{Model Reduction by ODENet}
Due to the existence of time-scale separation, it is possible to make a simplification of the reaction system in (\ref{mae4}). Classically, this is done by analytical methods, like Quasi Steady-State Assumption and Partial Equilibrium  Assumption \cite{Segel1989The, Huang2015Partial}. Here, we are going to show how the simplification procedure can be carried out automatically by ODENet.

\begin{table}[H]
	\setlength\extrarowheight{5pt}
	\setlength{\tabcolsep}{8pt}
	\setlength{\arraycolsep}{2pt}
    \centering
	\begin{tabular}{|l|l|l|l|l|l|}
		\hline
		\begin{tabular}[c]{@{}l@{}}Pearson's coefficient\end{tabular} & $dc_{P}/dt$ & $dc_{M}/dt$ & $dc_{G}/dt$ & $dc_{GP}/dt$ & $dc_{GP_2}/dt$ \\ \hline
		$dc_{P}/dt$                                                               & 1           & 0.9681      & 0.5020      & 0.4228       & 0.4215         \\ \hline
		$dc_{M}/dt$                                                               &             & 1           & 0.2743      & 0.1832       & 0.1820         \\ \hline
		$dc_{G}/dt$                                                               &             &             & 1           & 0.9892       & 0.9812         \\ \hline
		$dc_{GP}/dt$                                                              &             &             &             & 1            & 0.9956         \\ \hline
		$dc_{GP_2}/dt$                                                            &             &             &             &              & 1              \\ \hline
	\end{tabular}
	\caption{Pearson's correlation coefficients among time derivatives of five concentration variables in (\ref{mae4}).}
	\label{tab.correlation}
\end{table}

At the first step, with the help of traditional classification algorithms, like the correlation analysis based on the Pearson's coefficient between concentration
derivatives (see Table. \ref{tab.correlation}), the fast and slow variables can be easily separated into two groups. Inspired by the classical results of Michaelis-Menton kinetics, we suppose three fast variables $c_G, c_{GP}, c_{GP_2}$ (see Fig. \ref{fig.qssa}b-\ref{fig.qssa}e) are characterized by fractional functions, whose numerator and denominator are polynomials of $c_P$ (up to the second-order in the current study). In contrast, $c_M$ does not appear in the fractional functions, since the last three formulas in (\ref{mae4}) contain no terms of $c_M$. Consequently, the simplified model we are seeking for is given by
\begin{equation}
\left\{\begin{array}{l}
\vspace{2ex}
\displaystyle\frac{d}{d t} c_{P}=k_{s} c_{M}\underline{-k_{1} c_{P} H\left(c_{G}\right)+k_{-1} H\left(c_{G P}\right)-k_{2} c_{P} H\left(c_{G P}\right)+k_{-2} H\left(c_{G P_{2}}\right)}, \\
\vspace{2ex}
\displaystyle\frac{d}{d t} c_{M}=-k_{d M} c_{M}+k_{0} H\left(c_{G}\right)+k_{0} H\left(c_{G P}\right),\\
\vspace{2ex}
\displaystyle H\left(c_{G}\right)=\displaystyle\frac{\Omega_{1}}{\Omega}, H\left(c_{G P}\right)=\displaystyle\frac{\Omega_{2}}{\Omega}, H\left(c_{G P_{2}}\right)=\displaystyle\frac{\Omega_{3}}{\Omega},
\end{array}\right.
\label{qssa2}
\end{equation}
where $\Omega=\beta_{1}+\beta_{2} c_{p}+\beta_{3} c_{P}^{2}$, $\Omega_{1}=\alpha_{11}+\alpha_{21} c_{P}+\alpha_{31} c_{P}^{2}$, $\Omega_{2}=\alpha_{12}+\alpha_{22} c_{P}+\alpha_{32} c_{P}^{2}$, $\Omega_{3}=\alpha_{13}+\alpha_{23} c_{P}+\alpha_{33} c_{P}^{2}$.

\begin{table}[H]
	\setlength\extrarowheight{3pt}
	\setlength{\tabcolsep}{15pt}
    \centering
	\begin{tabular}{ccccccc}
		\Xhline{1.2pt}
		\hline
		\multicolumn{7}{c}{Parameters}                                                                                                \\ \hline
		& $\beta_{1}$    & $\beta_{2}$     & $\beta_{3}$  & $\alpha_{11}$      & $\alpha_{21}$ & $\alpha_{31}$    \\ \hline
		QSSA       & 0.53            & 0.67             & 1             & 0.0159             & 0             & 0                \\ \hline
		ODENet & 0.54 & 0.66   & 1   & 0.0163 & 0             & 0                \\ \hline
		Relative error   & 1.89\%           & 1.49\%           & $\sim$        & 2.52\%             & $\sim$        & $\sim$           \\ \hline
		& $\alpha_{12}$   & $\alpha_{22}$    & $\alpha_{32}$ & $\alpha_{13}$      & $\alpha_{23}$ & $\alpha_{33}$    \\ \hline
		QSSA       & 0               & 0.0201           & 0             & 0                  & 0             & 0.03             \\ \hline
		ODENet & 0               & 0.020 & 0             & 0                  & 0             & 0.03 \\ \hline
		Relative error   & $\sim$          & 0.50\%           & $\sim$        & $\sim$             & $\sim$        & 0\%              \\ \hline
		\Xhline{1.2pt}
	\end{tabular}
	\caption{Comparison on the learned parameters for (\ref{qssa2}) by ODENet with those by QSSA. All values are normalized by $\beta_{3}$.}
	\label{qssa_vc}
\end{table}

$\beta_1,\cdots,\beta_3$ and $\alpha_{11},\cdots,\alpha_{33}$ are twelve free parameters to be specified. As summarized in Table. \ref{qssa_vc}, the simplified model learned by ODENet is very close to that by QSSA (see next section). In particular, terms of $\alpha_{21}c_{P}$, $\alpha_{31} c_{P}^{2}$, $\alpha_{12}$, $\alpha_{32} c_{P}^{2}$, $\alpha_{13}$ and $\alpha_{23} c_{P}$ are removed by sparse identification during the learning procedure. A major difference between two simplification methods lies in the extra four underlined terms on the right-hand side of the first formula in (\ref{qssa2}). In QSSA, these four terms are exactly cancelled by each other. While during the simplification procedure aided by ODENet, we can only conclude that their sum is quite small instead of exactly zero (see SI).

\subsubsection{Comparison with QSSA}
Our above ODENet aided model reduction is consistent with the classical quasi steady-state approximation. Since $G$, $GP$, $GP_2$ are considered as the fast intermediates, in contrast to the slow species $P$ and $M$, a direct application of QSSA to (\ref{mae4}) leads to
\begin{equation}\label{algeq}
\begin{split}
&\frac{c_G}{c_{total}}=\frac{K_3}{\Omega}, \quad
\frac{c_{GP}}{c_{total}}=\frac{K_2 c_P}{\Omega},\quad
\frac{c_{GP_2}}{c_{total}}=\frac{{c_P}^2}{\Omega},
\end{split}
\end{equation}
where $\Omega = K_3 + K_2c_P + {c_P}^2$, $K_1=k_{-1}/k_1$, $K_2=k_{-2}/k_2$, $K_3=k_{-1}k_{-2}/(k_1k_2)$, and $c_{total}=c_G+c_{GP}+c_{GP_2}$ is a constant.

The corresponding reduced equations are
\begin{equation}\label{QSSAreduced}
\left\{
\begin{split}
&\frac{d}{dt}c_P= k_s c_{M},\\
&\frac{d}{dt}c_{M}=- k_{dM} c_{M} + k_0({K_3+ K_2 c_P})c_{total}/{\Omega},\\
&
\frac{c_G}{c_{total}}=\frac{K_3}{\Omega}, \quad
\frac{c_{GP}}{c_{total}}=\frac{K_2 c_P}{\Omega},\quad
\frac{c_{GP_2}}{c_{total}}=\frac{{c_P}^2}{\Omega},
\end{split}
\right.
\end{equation}
which has been used to evaluate the performance of our ODENet aided model reduction.

\section{Conclusion}

Nowadays, various machine learning algorithms, like deep learning and reinforcement learning, have found their applications in diverse fields with great success. While in the field of chemical reactions, related studies begin to emerge, yet are still quite few. In the current paper, through two concrete biochemical examples, the single proliferative compartment model and a gene network with autoregulatory negative feedback, we present our key ideas on how machine learning and multiscale modeling can help each other during the study of chemical reactions. And, as we believe, an effective integration of two approaches will be crucial for the success of related studies in this direction.

Potential generalizations of our current work include but are not limited to:

(1) \textbf{The spacial heterogeneity of chemical reactions.} In the current study, all reactions are assumed to proceed under well-mixed conditions, which means we can adopt a relatively simple ODE-based description. However, it is well-known the spacial heterogeneity can produce far more complicated and also interesting phenomena \cite{Janos1989Mathematical}, like the Turing pattern, phase separation, active matter, \textit{etc.} So how to generalize our results to PDEs would be of general interest. Recently, PDE-based machine learning algorithms \cite{Rudy2017Data, Raissi2017Numerical} shed light on this aspect.

(2) \textbf{Bistability, oscillation, bifurcation of chemical reactions.} Even restricted to ODEs, a chemical reaction system can possess very complex dynamical behaviors, like bistability, oscillation, bifurcation, blow-up, \textit{etc.}, than one can imagine \cite{Qian2009Stochastic, Bishop2010Stochatic}. In the presence of noise, the situation becomes even more complicated. The high-nonlinearity of chemical reactions puts forward great challenges to our ODENet-based model derivation and model reduction.

(3) \textbf{Extension to other model reduction methods.} Here we test the possibility and accuracy of ODENet aided model reduction with respect to the QSSA method. Extension of our ideas to partial equilibrium approximation \cite{Huang2015Partial}, maximum entropy principle \cite{jaynes1957information}, maximal likelihood estimation \cite{Harrell2015Overview}, as well as other statistics or probability based approximations would be worthy of further studies.

\section*{Acknowledgment}
This work was supported by the National Science Foundation of China (Grant No. 21877070 and 11871299), the Startup Research Funding of Minjiang University (mjy19033), and the Special Project of COVID-19 Epidemic Prevention and Control by Fuzhou Science and Technology Bureau (2020-XG-002). The authors would like to thank the helpful discussions from Dr. Pipi Hu.

\section*{AIP PUBLISHING DATA SHARING POLICY}

All the data in this paper which support the findings of this study are available from the corresponding author upon reasonable request.

\bibliography{multiscale}
\end{document}